\documentclass[twocolumn,floatfix,aps]{revtex4-2}
\usepackage{graphicx}
\usepackage{amsmath}
\usepackage{amssymb}
\usepackage{bm}
\usepackage{hyperref}

\usepackage{color}

\begin{document}

\title{Josephson effect in a junction coupled to an electron reservoir}
\author{C. W. J. Beenakker}
\affiliation{Instituut-Lorentz, Universiteit Leiden, P.O. Box 9506, 2300 RA Leiden, The Netherlands}
\date{April 2024}

\begin{abstract}
We extend the scattering theory of the Josephson effect to include a coupling of the Josephson junction to a gapless electron reservoir in the normal state. By opening up the system with a quasiparticle escape rate $1/\tau$, the supercurrent carried at zero temperature by an Andreev level at energy $\varepsilon_{\rm A}$ is reduced by a factor $(2/\pi)\arctan(2\varepsilon_{\rm A}\tau/\hbar)$ . We make contact with recent work on ``non-Hermitian Josephson junctions'', by comparing this result to different proposed generalizations of the Josephson effect to non-Hermitian Hamiltonians. 
\end{abstract}
\maketitle

\begin{figure}[tb]
\centerline{\includegraphics[width=1\linewidth]{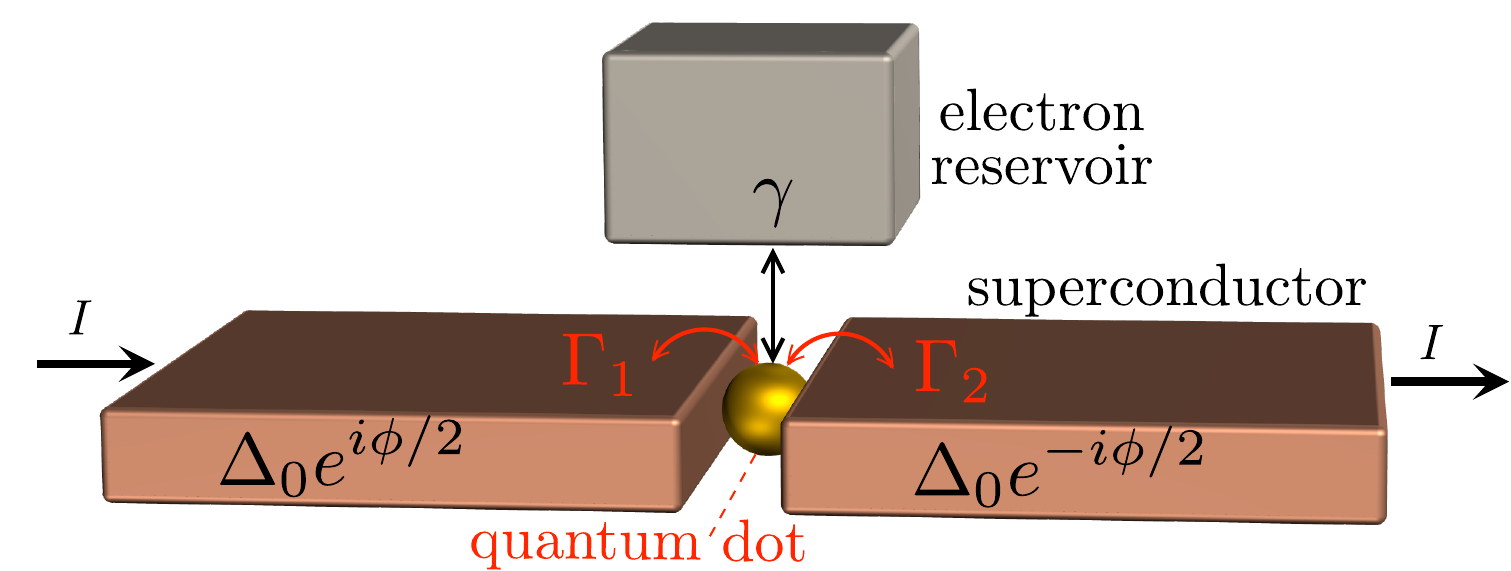}}
\caption{Josephson junction, formed by a quantum dot that is tunnel coupled with rates $\Gamma_1$ and $\Gamma_2$ to two superconductors (gap $\Delta_0$, phase difference $\phi$). The quantum dot has an additional weak coupling (rate $\gamma$) to a gapless electron reservoir in the normal state. We seek the $\gamma$-dependence of the current-phase relationship $I(\phi)$.
}
\label{fig_diagram}
\end{figure}

\textit{Introduction ---}
The thermodynamic properties of a quantum dot are affected by the coupling to a superconductor. The Josephson effect is a striking example, a current will flow through the quantum dot in equilibrium if the quantum dot forms a weak link between two superconductors \cite{Gol04}. This supercurrent $I(\phi)$ depends periodically on the phase difference $\phi$ of the pair potential in the superconductors. A dissipative electromagnetic environment may degrade the supercurrent via phase fluctuations \cite{Ave98,Pek01,Mak04}. Here we investigate an altogether different decay mechanism, the coupling of the quantum dot to a gapless electron reservoir (see Fig.\ \ref{fig_diagram}).

The reservoir is a source of dephasing because quasiparticles that enter it from the quantum dot return without any phase coherence. Such a dephasing mechanism for persistent currents was introduced by B\"{u}ttiker \cite{But85}, and applied to superconducting circuits by several authors \cite{Wee91,Cha97,Mor00,Gra00,Bel01,Bel03,Ber09}. Our main advance is that we obtain closed-form expressions for the supercurrent-phase relationship, for the case of a spatially uniform coupling rate to the electron reservoir. 

Because the effective Hamiltonian of the quantum dot becomes non-Hermitian when we open it up to the reservoir, such a system is a simple example of a ``non-Hermitian Josephson junction'', a topic of current interest \cite{Kor23,Li23,Cay23,She24}. Our explicit expressions support Ref.\ \onlinecite{She24} in the debate over the proper generalization of the current-phase relationship $I\propto dE/d\phi$ to complex eigenvalues $E$ of the effective Hamiltonian.

\textit{Closed system ---}
In a scattering formulation \cite{Bee91}, the dependence of the density of states $\rho$ of the Josephson junction on the superconducting phase difference $\phi$ is given by the determinantal expression 
\begin{align}
\rho(\varepsilon)={}&-\frac{1}{\pi}\operatorname{Im}\frac{d}{d\varepsilon}\ln\operatorname{det}\bigl[1-R_{\rm A}(\varepsilon+i0^+)S_{\rm N}(\varepsilon+i0^+)\bigr]\nonumber\\
&+\rho_{0}(\varepsilon),\label{rhodef}
\end{align}
with $\rho_0$ the $\phi$-independent density of states of the junction when it is decoupled from the superconductors. The determinant contains the product of the Andreev reflection matrix $R_{\rm A}$ from the superconductors and the scattering matrix $S_{\rm N}$ of the junction in the normal state. The energy $\varepsilon>0$ is the excitation energy of Bogoliubov quasiparticles (electron-hole superpositions). The pair potential in the two superconductors, to the left and to the right of the junction, has amplitude $\Delta_0$ and phase $\pm\phi/2$.

The electron and hole degree of freedom introduces a block structure in the scattering matrices. The matrix $S_{\rm N}$ is block diagonal,
\begin{equation}
S_{\rm N}(\varepsilon)=\begin{pmatrix}
s_0(\varepsilon)&0\\
0&s_0^\ast(-\varepsilon)
\end{pmatrix},\label{SNdef}
\end{equation}
the electron and hole blocks are uncoupled and related by particle-hole symmetry.

The Andreev reflection matrix has the block structure
\begin{align}
&R_{\rm A}(\varepsilon)=i\alpha(\varepsilon)\begin{pmatrix}
0&r_{\rm A}\\
r^\ast_{\rm A}&0
\end{pmatrix},\;\;
r_{\rm A}=
\begin{pmatrix}
e^{i\phi/2}\sigma_y&0\\
0&e^{-i\phi/2}\sigma_y
\end{pmatrix},\nonumber\\
&\alpha(\varepsilon)=e^{-i\arccos(\varepsilon/\Delta_0)}=\varepsilon/\Delta_0-i\sqrt{1-\varepsilon^2/\Delta_0^2}.
\label{RAdef}
\end{align}
The Pauli matrix $\sigma_y$ acts on the spin degree of freedom. The blocks $r^\ast_{\rm A}$ and $r_{\rm A}$ describe Andreev reflection from, respectively, electron to hole and hole to electron, in opposite spin bands. Andreev reflection happens with unit probability for energies $\varepsilon<\Delta_0$; any normal reflection at the normal--superconductor (NS) interface is incorporated into $S_{\rm N}$.

Substitution of Eqs.\ \eqref{SNdef} and \eqref{RAdef} into Eq.\ \eqref{rhodef} gives the determinant \cite{Bee91}
\begin{align}
&\rho(\varepsilon)=-\frac{1}{\pi}\operatorname{Im}\frac{d}{d\varepsilon}\ln\operatorname{det}\bigl[1-M(\varepsilon+i0^+)\bigr]+\rho_0(\varepsilon),\label{rhoalphas0}\\
&M(\varepsilon)=-\alpha(\varepsilon)^2 r_{\rm A}^\ast s_0^\ast(-\varepsilon)  r_{\rm A}^{\vphantom{\ast}}s_0(\varepsilon) .\label{Mdef}
\end{align}
The density of states determines the free energy $F$ of the Josephson junction at temperature $T$ \cite{Bar69}
\begin{align}
F=-k_{\rm B}T\int_0^\infty d\varepsilon\,\rho(\varepsilon)\ln\bigl[2\cosh(\varepsilon/2k_{\rm B}T)\bigr].\label{omegadef}
\end{align}
The supercurrent through the Josephson junction then follows from the relation
\begin{equation}
I=\frac{2e}{\hbar}\frac{dF}{d\phi}.\label{IdFdphi}
\end{equation}

It is convenient to extend the integration range in Eq.\ \eqref{omegadef} to negative $\varepsilon$ (so we no longer need to extract the imaginary part) and then to perform a partial integration,
\begin{align}
F={}&\frac{i}{4\pi}\int_{-\infty}^\infty d\varepsilon\,\tanh(\varepsilon/2k_{\rm B}T)\ln\operatorname{det}\bigl[1-M(\varepsilon+i0^+)\bigr]\nonumber\\
&+F_0,
\end{align}
with $F_0$ the $\phi$-independent contribution to the free energy.

\textit{Opening up the system ---}
We open up the Josephson junction by weakly coupling it to a gapless electron reservoir in the normal state. Quasiparticles enter the reservoir at a rate $1/\tau\equiv 2\gamma/\hbar$, assumed to be spatially uniform in the junction region. This can be modeled by coupling to the reservoir via a spatially extended tunnel barrier \cite{Zir93,Bro97a}. The reservoir is closed, it does not drain any current, but particles that enter it return to the junction without any phase coherence, so they no longer contribute to the supercurrent. 

The full scattering matrix of the normal region, including the scattering channels to the electron reservoir, is unitary, with $S_{\rm N}(\varepsilon+i\gamma)$ a sub-unitary submatrix for the scattering channels to the superconductors. Similarly, the Andreev reflection matrix $R_{\rm A}(\varepsilon+i\gamma)$ is now sub-unitary also for $\varepsilon<\Delta_0$.

Instead of Eq.\ \eqref{rhoalphas0} we have for the density of states the expression
\begin{equation}
\rho(\varepsilon)=-\frac{1}{\pi}\operatorname{Im}\frac{d}{d\varepsilon}\ln\operatorname{det}\bigl[1-M(\varepsilon+i\gamma)\bigr]+\rho_{0}(\varepsilon).\label{rhodefvarphi}
\end{equation}
The term $\rho_0$ now also contains the $\phi$-independent contributions from the electron reservoir. The free energy then follows from
\begin{align}
F={}&\frac{i}{4\pi}\int_{-\infty}^\infty d\varepsilon\,\tanh(\varepsilon/2k_{\rm B}T)\ln\operatorname{det}\bigl[1-M(\varepsilon+i\gamma)\bigr]\nonumber\\
&+F_0.\label{Fgamma}
\end{align}
We discuss three applications of this general formula.

\textit{Quantum dot Josephson junction ---}
As a first application we consider the case that the two superconductors are coupled via a quantum dot containing a single resonant level, at energy $\varepsilon_{\rm R}$ relative to the Fermi level. We ignore Coulomb blockade effects, which will be less significant in the open system.

The electronic scattering matrix of the quantum dot has the form
\begin{equation}
s_0(\varepsilon)=\begin{pmatrix}
1&0\\
0&1
\end{pmatrix}-\frac{i}{\varepsilon-\varepsilon_{\rm R}+\tfrac{1}{2}i\Gamma}\begin{pmatrix}
\Gamma_1&\sqrt{\Gamma_1\Gamma_2}\\
\sqrt{\Gamma_1\Gamma_2}&\Gamma_2
\end{pmatrix},
\end{equation}
with $\Gamma_1,\Gamma_2$ the tunnel rates into the left and right superconductor, and $\Gamma=\Gamma_1+\Gamma_2$. The normal-state transmission probability through the quantum dot at the Fermi level ($\varepsilon=0$) is given by the Breit-Wigner formula \cite{Bre36}
\begin{equation}
T_{\rm BW}=\frac{\Gamma_1\Gamma_2}{\varepsilon_{\rm R}^2+\tfrac{1}{4}\Gamma^2}.
\end{equation}

In the weak-coupling, near-resonant regime, when $\varepsilon_{\rm R},\Gamma\ll\Delta_0$, the energy dependence of $R_{\rm A}$ can be neglected and we may substitute $\alpha(\varepsilon)\rightarrow\alpha(0)= -i$. The determinant \eqref{rhoalphas0} evaluates to
\begin{equation}
\ln\det[1-M(\varepsilon)]=g\ln\left[1-(\varepsilon_{\rm A}/\varepsilon)^2\right],\label{detMqd}
\end{equation}
plus $\phi$-independent terms. The prefactor $g$ accounts for spin and possibly other degeneracies. The energy 
\begin{equation}
\varepsilon_{\rm A}=\Delta_{\rm eff}\sqrt{1-T_{\rm BW} \sin^2( \phi/2)},\;\;\Delta_{\rm eff}=\sqrt{ \varepsilon_{\rm R}^2 +\tfrac{1}{4}\Gamma^2},\label{eAdef}
\end{equation}
is the energy of the Andreev level in the closed system \cite{Bee92,Dev97}. There is no dependence on $\Delta_0$ in the weak-coupling regime.

\begin{figure}[tb]
\centerline{\includegraphics[width=0.8\linewidth]{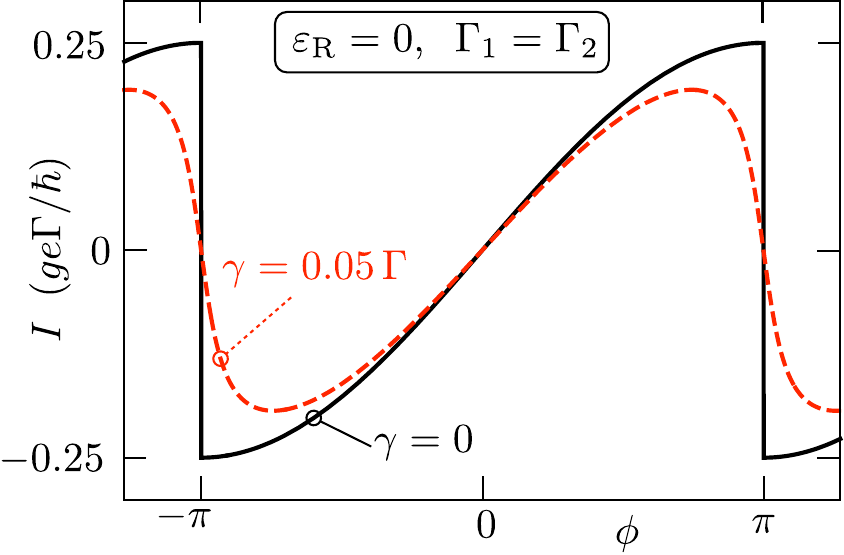}}
\caption{Zero-temperature current-phase relation of the quantum dot Josephson junction of Fig.\ \ref{fig_diagram}, with and without the coupling to the electron reservoir. The plot is computed from Eq.\ \eqref{Iqdtis0} for the case $T_{\rm BW}=1$ of unit transmission probability at the Fermi level through the quantum dot.
}
\label{fig_qd}
\end{figure}

The free energy \eqref{Fgamma} becomes
\begin{equation}
F=\frac{gi}{4\pi}\int_{-\infty}^\infty d\varepsilon\tanh(\varepsilon/2k_{\rm B}T)\ln\left[1-\frac{\varepsilon_{\rm A}^2}{(\varepsilon+i\gamma)^2}\right]+F_0.
\end{equation}
The resulting supercurrent is
\begin{align}
I={}&I_0\int_{-\infty}^\infty \frac{d\varepsilon}{i\pi}\frac{\varepsilon_{\rm A}\tanh(\varepsilon/2k_{\rm B}T)}{(\varepsilon_{\rm A})^2-(\varepsilon+i\gamma)^2}\nonumber\\
={}&I_0\frac{2}{\pi}\operatorname{Im} \psi \left(\frac{1}{2}+\frac{i \varepsilon_{\rm A}+\gamma}{2 \pi k_{\rm B}T }\right),\label{IfiniteT}
\end{align}
with $\psi(x)$ the digamma function and
\begin{equation}
I_0=-\frac{ge}{\hbar}\frac{d\varepsilon_{\rm A}}{d\phi}\label{I0qdzeroT}
\end{equation}
the zero-temperature supercurrent of the closed system. The $T\rightarrow 0$ limit of Eq.\ \eqref{IfiniteT} (plotted in Fig.\ \ref{fig_qd}) is
\begin{equation}
\lim_{T\rightarrow 0}I=I_0\frac{2}{\pi}\arctan(\varepsilon_{\rm A}/\gamma),\label{Iqdtis0}
\end{equation}
so opening up the system to an electron reservoir at $T=0$ reduces the supercurrent by a factor $(2/\pi)\arctan(2\varepsilon_{\rm A}\tau/\hbar)$.

\textit{Point contact Josephson junction ---}
The second application is a point contact junction of length $L$ short compared to the superconducting coherence length $\xi_0=\hbar v_{\rm F}/\Delta_0$. In this short-junction regime the energy dependence of $S_{\rm N}$ can be neglected relative to the energy dependence of $R_{\rm A}$, and we may evaluate $S_{\rm N}$ at the Fermi level. We assume that time reversal symmetry is preserved, hence $\sigma_y S^\ast_{\rm N}\sigma_y=S_{\rm N}^\dagger$. 

The Andreev levels in the closed system then depend on the phase difference according to \cite{Bee91}
\begin{equation}
\varepsilon_n=\Delta_0\sqrt{1-T_n\sin^2(\phi/2)},\;\;n=1,2,\ldots N,\label{epsnQPC}
\end{equation}
in terms of the mode-dependent transmission probabilities $T_n\in[0,1]$ in the normal state (eigenvalues of the transmission matrix product $tt^\dagger$). The number $N$ counts the number of propagating electron modes through the point contact. Eq.\ \eqref{epsnQPC} is analagous to the result \eqref{eAdef} for a quantum dot, with $\Delta_{\rm eff}$ replaced by $\Delta_0$ and $T_{\rm BW}$ replaced by $T_n$.

Instead of Eq.\ \eqref{detMqd} we now have
\begin{equation}
\ln\det[1-M(\varepsilon)]=g\sum_{n=1}^N\ln\left[1-(\varepsilon_{n}/\varepsilon)^2\right],
\end{equation}
and following the same steps as in the previous application we find the supercurrent in the open system,
\begin{align}
I={}&\sum_{n=1}^N I_n\frac{2}{\pi}\operatorname{Im} \psi \left(\frac{i \varepsilon_{n}+\gamma+\pi k_{\rm B}T }{2 \pi k_{\rm B}T }\right),\\
\lim_{T\rightarrow 0}I={}&\sum_{n=1}^N I_n\frac{2}{\pi}\arctan(\varepsilon_{n}/\gamma),
\end{align}
with $I_n=-(ge/\hbar)d\varepsilon_n/d\phi$ the zero-temperature supercurrent in the $n$-th mode of the closed system.

\textit{Long SNS junction ---}
For the third and final application we consider a superconductor--normal-metal--superconductor (SNS) junction of length $L\gg\xi_0$. Both states above and below $\Delta_0$ then contribute to the supercurrent. 

It is convenient to transform the slowly converging integration over energies of Eq.\ \eqref{Fgamma} into a more rapidly converging sum over Matsubara frequencies \cite{Bro97b},
\begin{equation}
\begin{split}
&F=-k_{\rm B}T\sum_{p=0}^\infty\ln\det[1-M(i\omega_p+i\gamma)],\\
&\omega_p=(2p+1)\pi k_{\rm B} T.
\end{split}\label{omegapvarphi}
\end{equation}
We will restrict ourselves for this application to zero temperature, when the sum can be replaced by an integral,
\begin{equation}
\lim_{T\rightarrow 0}F=-\frac{1}{2\pi}\int_{0}^\infty d\omega\,\ln\det[1-M(i\omega+i\gamma)].\label{OmegaMatsubara}
\end{equation}

We consider a ballistic single-mode junction. The electronic scattering matrix is
\begin{equation}
s_0(\varepsilon)=e^{-ik(\varepsilon) L}\begin{pmatrix}
0&1\\
1&0
\end{pmatrix}.
\end{equation}
We linearize the momentum near the Fermi energy, $k(\varepsilon)=k_{\rm F}+\varepsilon/\hbar v_{\rm F}$.

The determinant \eqref{Mdef} evaluates to
\begin{align}
&\ln\det[1-M(\varepsilon)]=\nonumber\\
&\quad =g\ln\left[1+\alpha^4 e^{4 i  \varepsilon L/\hbar v_{\rm F}}-2 \alpha^2 e^{2 i  \varepsilon L/\hbar v_{\rm F}} \cos \phi\right].
\end{align}
The sum over Matsubara frequencies decays on the scale of $\hbar v_{\rm F}/L$, which is much smaller than $\Delta_0$ in the long-junction regime. We may thus approximate $\alpha(i\omega)\approx\alpha(0)= -i$, when
\begin{equation}
\ln\det[1-M(i\omega)]=g\ln\left|1+e^{i\phi}e^{-2\omega L/\hbar v_{\rm F}}\right|,
\end{equation}
resulting in the zero-temperature free energy
\begin{equation}
F=\frac{g \hbar v_{\rm F}}{2\pi L}\operatorname{Re}\text{Li}_2\left(-e^{i \phi-2 \gamma L/\hbar v_{\rm F}}\right).
\end{equation}
The function $\text{Li}_2(x)$ is the dilogarithm.

\begin{figure}[tb]
\centerline{\includegraphics[width=0.8\linewidth]{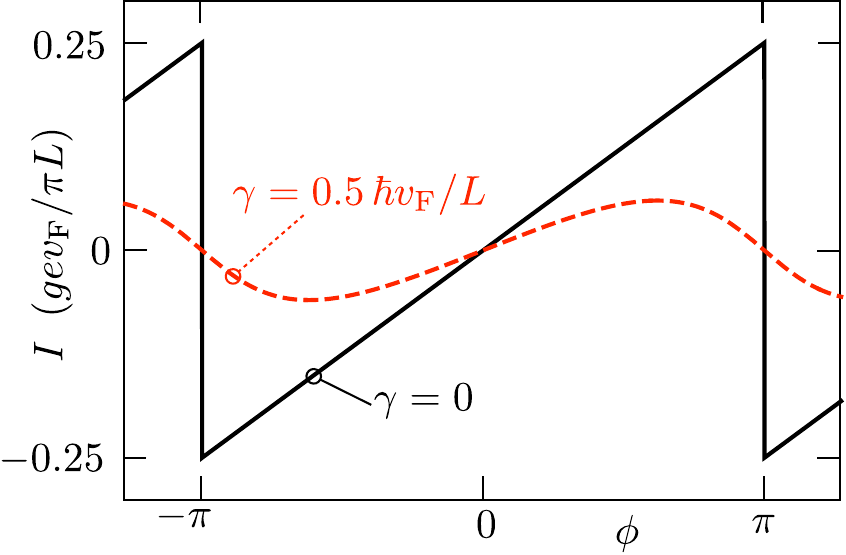}}
\caption{Same as Fig.\ \ref{fig_qd}, but now for a single-mode ballistic SNS junction in the long junction limit (length $L$ large compared to the superconducting coherence length $\xi_0=\hbar v_{\rm F}/\Delta_0$). The plot is computed from Eq.\ \eqref{isns0}.
}
\label{fig_sns}
\end{figure}

The corresponding supercurrent is
\begin{equation}
I=\frac{ge v_{\rm F}}{\pi L}\arctan\left[\frac{ \sin \phi}{e^{2\gamma L/\hbar v_{\rm F}}+ \cos \phi}\right],\label{isns0}
\end{equation}
see Fig.\ \ref{fig_sns}. As a check, in the limit $\gamma\rightarrow 0$ we recover the known sawtooth $\phi$-dependence \cite{Ish70,Bar72,Svi73},
\begin{equation}
\lim_{\gamma\rightarrow 0}I=\frac{gev_{\rm F}}{\pi L}\arctan\tan(\phi/2)=\frac{gev_{\rm F}}{\pi L}\phi,\;\;-\pi<\phi<\pi.
\end{equation}
Refs.\ \onlinecite{Wee91,Cha97} have qualitatively similar plots to Fig.\ \ref{fig_sns}, for different models of coupling to electron reservoirs that do not allow for a closed-form solution.

\textit{Connection to the non-Hermitian Josephson effect ---}
The coupling of the Josephson junction to the electron reservoir  pushes the poles of its scattering matrix into the lower half of the complex plane, down to $E=\pm\varepsilon_{\rm A}-i\gamma$. These poles can be considered as the complex eigenvalues of an effective non-Hermitian Hamiltonian ${\cal H}$. The open Josephson junction thus provides a physical realization of the non-Hermitian Josephson effect studied in Refs.\ \onlinecite{Kor23,Li23,Cay23,She24}. Different generalizations have been proposed for the relation $I\propto dE/d\phi$ when $E$ is complex. Let us compare these with our findings.

To be specific, we consider the weakly coupled quantum dot Josephson junction, with effective Hamiltonian \cite{Men09,Rec10,Ori21}
\begin{equation}
{\cal H}=\begin{pmatrix}
\varepsilon_{\rm R}-i\gamma&\tfrac{1}{2}e^{i\phi/2}\Gamma_1+\tfrac{1}{2}e^{-i\phi/2}\Gamma_2\\
\tfrac{1}{2}e^{-i\phi/2}\Gamma_1+\tfrac{1}{2}e^{i\phi/2}\Gamma_2&-\varepsilon_{\rm R}-i\gamma
\end{pmatrix}.\label{calH}
\end{equation}
One checks that the eigenvalues are $\pm\varepsilon_{\rm A}-i\gamma$, with $\varepsilon_{\rm A}$ given by Eq.\ \eqref{eAdef}.

Refs.\ \onlinecite{Li23} and \onlinecite{Cay23} argue that only the real part of $E$ contributes to the physical supercurrent, which in this case would imply no effect from the coupling to the reservoir. Shen, Lu, Lado, and Trif \cite{She24}, in a remarkable recent paper, give instead the zero-temperature relation
\begin{equation}
I=-\frac{2e}{\pi\hbar}\frac{d}{d\phi}\operatorname{Im}\operatorname{Tr}({\cal H}\ln{\cal H}),\label{HlnH}
\end{equation}
which reduces precisely to our result \eqref{I0qdzeroT} (with $g=2$) \cite{note1}.

\textit{Conclusion ---}
In summary, we have calculated how the coupling to a gapless electron reservoir in the normal state reduces the supercurrent through a Josephson junction. A simple answer is obtained for the model where the escape rate $1/\tau\equiv 2\gamma/\hbar$ of quasiparticles into the reservoir is spatially uniform. At zero temperature the reduction factor for a given Andreev level equals $(2/\pi)\arctan(2\varepsilon_{\rm A}\tau/\hbar)$. This applies to a weakly coupled quantum dot Josephson junction, or to a short point contact (length $L$ smaller than the coherence length $\xi_0$). A more complicated $\tau$-dependence is obtained in a long junction ($L\gg\xi_0$), when also states above the gap contribute to the supercurrent.

\textit{Acknowledgments ---}
I have benefited from discussions with A. R. Akhmerov, Yu.\ V. Nazarov, P.-X. Shen, and T. Vakhtel. This project has received funding from the European Research Council (ERC) under the European Union's Horizon 2020 research and innovation programme.


\begin{thebibliography}{99}
\bibitem{note1} To evaluate Eq.\ \eqref{HlnH} one can work in a basis where the Hamiltonian \eqref{calH} is diagonal, with eigenvalues $\pm\varepsilon_{\rm A}-i\gamma$, and note that $(d/d\phi)\operatorname{Tr}{\cal H}=0$ while $d{\cal H}/d\phi$ is real, so that $(d/d\phi)\operatorname{Im}\operatorname{Tr}({\cal H}\ln{\cal H})=\operatorname{Tr}(d{\cal H}/d\phi)\operatorname{Im}\ln{\cal H}=(d\varepsilon_{\rm A}/d\phi)\operatorname{Im}\ln\frac{\varepsilon_{\rm A}-i\gamma}{-\varepsilon_{\rm A}-i\gamma}=2(d\varepsilon_{\rm A}/d\phi)\arctan(\varepsilon_{\rm A}/\gamma).$ This is at zero temperature. We have checked that the finite-$T$ expression \eqref{IfiniteT} also agrees with Ref.\ \onlinecite{She24}.
\end{thebibliography}
\end{document}